\documentclass{moriond}

\def\Journal#1#2#3#4{{#1} {\bf #2}, #3 (#4)}

\def\NCA{\em Nuovo Cimento}

\def\PLB{{\em Phys. Lett.}  B}
\def\PRL{\em Phys. Rev. Lett.}
\def\PRD{{\em Phys. Rev.} D}

\def\PhysRept{{\em Phys. Rept.}}
\def\JHEP{{\em JHEP}}
\def\JINST{{\em JINST}}
\def\EPJC{{\em Eur. Phys. J.} C}
\def\JPR{{\em J. Phys. Radium}}
\def\PhysRev{{\em Phys. Rev.}}
\def\JETP{{\em J. Exp. Theor. Phys.}}
\def\PoS{{\em PoS}}
\def\Nature{{\em Nature}}

\usepackage{amsmath,amssymb,gensymb}

\newcommand{\Photo}%{\includegraphics[height=35mm]{mypicture}}

\newcommand{\beq}{\begin{equation}}
\newcommand{\eeq}{\end{equation}}

\newcommand{\GeV}{\,\text{GeV}}

\begin{document}
\vspace*{4cm}
\title{Dispersive approaches for the HVP and HLbL contributions to $\boldsymbol{(g-2)_\mu}$}

\author{M.~Hoferichter}

\address{Albert Einstein Center for Fundamental Physics, Institute for Theoretical Physics, University of Bern, Sidlerstrasse 5, 3012 Bern, Switzerland}

\maketitle\abstracts{Calculations based on the analytic properties of the required matrix elements allow for a wide range of applications constraining the hadronic contributions to the anomalous magnetic moment of the muon $a_\mu=(g-2)_\mu/2$, both hadronic vacuum polarization (HVP) and hadronic light-by-light (HLbL) scattering. Here, we discuss such recent applications, including analyticity constraints on hadronic cross sections, radiative corrections, and isospin-breaking effects.}

\section{Introduction}

The HVP contribution to $a_\mu$ has been estimated with dispersive techniques as early as in Refs.~\cite{Bouchiat:1961lbg,Brodsky:1967sr}, relating the leading-order (LO) contribution to the total cross section for $e^+e^-\to\text{hadrons}$ via 
\beq
 a_\mu^\text{HVP,\,LO}=\bigg(\frac{\alpha m_\mu}{3\pi}\bigg)^2\int_{s_\text{thr}}^\infty ds \frac{\hat K(s)}{s^2}R_\text{had}(s),\qquad R_\text{had}(s)=\frac{3s}{4\pi\alpha^2}\sigma(e^+e^-\to\text{hadrons}+\gamma),
 \eeq
where $\hat K(s)$ is a known kernel function and the cross section is defined including radiative channels, in such a way that the integration starts at $s_\text{thr}=M_{\pi^0}^2$ due to $e^+e^-\to\pi^0\gamma$. The HVP evaluation of Ref.~\cite{Aoyama:2020ynm} is based on direct integration of the cross-section data~\cite{Davier:2017zfy,Keshavarzi:2018mgv}, but at the same time further constraints on the $2\pi$ and $3\pi$ cross sections~\cite{Colangelo:2018mtw,Hoferichter:2019mqg} from analyticity, unitarity, and crossing symmetry were taken into account.  Such constraints arise because the leading hadronic channels are described by reasonably simple matrix elements---electromagnetic form factors for $\pi^+\pi^-$ and $\bar K K$, $\gamma^*\to 3\pi$ for the $3\pi$ channel---which are again strongly constrained by their analytic properties. Besides refinements in the $2\pi$ and $3\pi$ cross sections~\cite{Colangelo:2020lcg,Colangelo:2022prz,Stoffer:2023gba,Hoferichter:2023bjm}, also dispersive evaluations for $\pi^0\gamma$ and $\bar K K$ have become available in recent years~\cite{Hoid:2020xjs,Stamen:2022uqh}. The motivation for dispersive studies of the hadronic cross sections includes (i) cross checks on the experimental data sets, (ii) improved precision in kinematic ranges where data are scarce, (iii) correlations with other low-energy observables, (iv) structure-dependent radiative corrections, and (v) comparison to lattice QCD.  For HLbL scattering, a dispersive approach becomes much more complicated because in contrast to the two-point function $\Pi_{\mu\nu}=\langle0|T\{j_\mu j_\nu\}|0\rangle$ that describes HVP, the hadronic four-point function $\Pi_{\mu\nu\lambda\sigma}=\langle0|T\{j_\mu j_\nu j_\lambda j_\sigma\}|0\rangle$ needs to be reconstructed~\cite{Colangelo:2014dfa,Colangelo:2014pva}.  

\section{Hadronic vacuum polarization}
\label{sec:HVP}

\subsection{Analyticity constraints on hadronic cross sections}

As the arguably simplest case, we discuss the example of the electromagnetic form factor of the pion $F_\pi^V(s)$, for which a dispersive approach leads to a decomposition of the form 
\beq
F_\pi^V(s) = \underbrace{\Omega_1^1(s)}_\text{elastic $\pi\pi$ scattering}\times \underbrace{G_\omega(s)}_\text{isospin-breaking $3\pi$ cut}\times \underbrace{G_\text{in}(s)}_\text{inelastic effects: $4\pi$, \ldots}.
\eeq
Here, $\Omega_1^1(s)$ refers to the Omn\`es factor~\cite{Omnes:1958hv}, incorporating the dominant effect of $2\pi$ intermediate states in terms of the $\pi\pi$ $P$-wave phase shift. This phase shift is further strongly constrained by $\pi\pi$ Roy equations~\cite{Roy:1971tc}, leading to a parameterization in terms of phase-shift values at just two energies~\cite{Caprini:2011ky}. $G_\omega(s)$ parameterizes the $3\pi$ cut in terms of resonance parameters and residue of the $\omega(782)$, including higher-order isospin-breaking corrections from $\rho$--$\omega$ mixing via radiative channels~\cite{Colangelo:2022prz}. Finally, $G_\text{in}(s)$ parameterizes inelastic effects, primarily $4\pi$ intermediate states, in terms of a conformal polynomial. In this way, correlations of the $e^+e^-\to2\pi$ cross section arise with $\pi\pi$ scattering phase shifts, $\omega$ resonance parameters, the pion charge radius, and potentially other low-energy observables~\cite{Colangelo:2020lcg,Heuser:2024biq}. A similar parameterization can be used to predict the quark-mass dependence of the resulting HVP contribution~\cite{Colangelo:2021moe,Niehus:2020gmf}.

\begin{figure}[t] 
	\centering\small
	\includegraphics[width=0.495\linewidth]{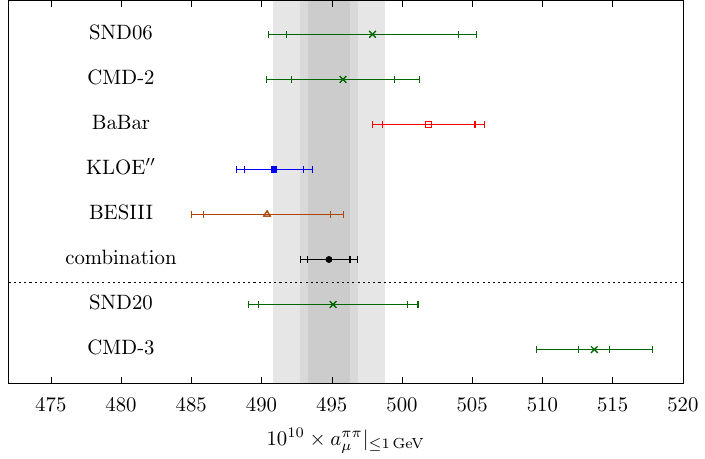}
	\includegraphics[width=0.495\linewidth]{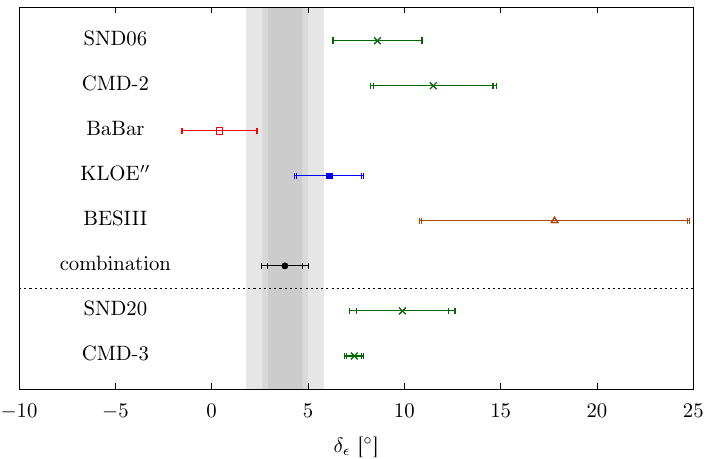}
	\caption{Results for the $2\pi$ contribution to HVP in the energy range 
	below $1\GeV$ (left) and for the phase of the $\rho$--$\omega$ mixing parameter, $\delta_\epsilon$ (right). In both cases, the smaller error refers to $\chi^2$-inflated fit uncertainties, the larger one to the total uncertainty including the systematics of the dispersive representation. 
	For the combined fit, the third gray band, in addition, includes an estimate of the BaBar--KLOE tension.}
	\label{fig:Plots}
\end{figure}

The application to the $2\pi$ contribution is shown in Fig.~\ref{fig:Plots}, see Ref.~\cite{Stoffer:2023gba}. First, the use of the dispersive representation allows us to compare all data sets in the entire low-energy domain, not just in the range where data are available, and quantify the tensions accordingly (see left panel in Fig.~\ref{fig:Plots}). We find that throughout the low-energy region, the tension between CMD-3~\cite{CMD-3:2023alj} and the previous average~\cite{Achasov:2006vp,CMD-2:2006gxt,BaBar:2012bdw,BESIII:2015equ,KLOE-2:2017fda} evaluates to $\gtrsim 5\sigma$, thus compounding the tension visible only in a restricted energy range~\cite{CMD-3:2023alj}. Interestingly, we observe that the fit quality to the CMD-3 data is good, in contrast to SND-20~\cite{SND:2020nwa}, with a change in central value that is mainly driven by the inelastic contributions, while the resulting values for the $\pi\pi$ phase shifts are only mildly affected. Since, at the same time, the theory error is dominated by the truncation of the conformal polynomial, the dispersive analysis thus suggests that further constraints that might help clarify the situation could be derived by studying correlations with the inelastic channels, i.e., data on $e^+e^-\to 4\pi,\pi\omega,\ldots$ Another interesting observation is illustrated in the right panel in Fig.~\ref{fig:Plots}, related to the $\rho$--$\omega$ mixing parameter, which can be identified with the residue of the $\omega$  contribution and ultimately determines the strength of $G_\omega(s)$. Both the modulus, for which there is reasonable agreement among the experiments, and the phase 
$\delta_\epsilon$ are observable. The latter vanishes in the isospin limit, but can take a non-zero value due to radiative intermediate states $\rho\to \pi^0\gamma, \eta\gamma, \pi\pi\gamma,\ldots \to \omega$, based on which one finds a narrow-resonance estimate $\delta_\epsilon=3.5(1.0)\degree$~\cite{Colangelo:2022prz}. While the combined fit agrees well with this estimate, the considerable spread among the individual experiments points to further tensions besides the overall size of the cross section and, moreover, 
the pattern of the spread is not correlated with the one for the HVP integral. 

\subsection{Radiative corrections}

Radiative corrections to the ($C$-odd) forward--backward asymmetry for $e^+e^-\to\pi^+\pi^-$ involve box diagrams that are sensitive to pion structure. This was first  observed in a model context in Ref.~\cite{Ignatov:2022iou}, but using a 
dispersive representation for the pion form factor allows one to evaluate such structure-dependent radiative corrections in a theoretically robust framework that avoids unphysical imaginary parts~\cite{Colangelo:2022lzg}. The measurement of the asymmetry by CMD-3~\cite{CMD-3:2023alj} showed that the infrared enhancement makes such corrections relevant, leading to the question whether similar diagrams could become important for initial-state-radiation experiments, for which they can contribute to the ($C$-even) cross section~\cite{Abbiendi:2022liz}.

\subsection{Isospin breaking}

Having obtained a detailed understanding of the hadronic cross sections for the dominant channels, we estimated the resulting isospin-breaking effects summing up the radiative channels $\pi^0\gamma$, $\eta\gamma$, $\omega(\to\pi^0\gamma)\pi^0$, final-state radiation in $2\pi$, $3\pi$, $K^+K^-$, $\rho$--$\omega$ mixing in $2\pi$, $3\pi$, as well as mass differences in $2\pi$, $\bar K K$~\cite{Hoferichter:2023sli,Hoferichter:2022iqe}. A number of individually large effects cancel to a large extent, leaving an overall relatively small correction that agrees reasonably well with lattice-QCD calculations~\cite{Borsanyi:2020mff,RBC:2018dos}. Since, moreover, the lattice-QCD result for HVP would even become larger if such phenomenological estimates for isospin breaking were adopted instead of the dedicated lattice-QCD calculation,  isospin-breaking corrections are very unlikely to be the reason for the tension between lattice QCD and phenomenology, including the intermediate window~\cite{Borsanyi:2020mff,Colangelo:2022vok,Ce:2022kxy,ExtendedTwistedMass:2022jpw,FermilabLatticeHPQCD:2023jof,RBC:2023pvn}.

\section{Hadronic light-by-light scattering}
\label{sec:HLbL}

To match the projected final result from the Fermilab experiment~\cite{Muong-2:2023cdq,Colangelo:2022jxc}, the precision of the HLbL contribution needs to be improved by a factor $2$ compared to Ref.~\cite{Aoyama:2020ynm}.
There has been recent progress
in lattice QCD towards this goal~\cite{Chao:2021tvp,Blum:2023vlm}, but also dispersive methods are well poised to achieve the required precision: with the dominant intermediate states already evaluated~\cite{Colangelo:2017qdm,Hoferichter:2018dmo,Danilkin:2021icn}, ongoing work includes dispersive evaluations of $\eta$, $\eta'$ poles~\cite{Holz:2015tcg,Holz:2022hwz}, axial-vector and tensor contributions~\cite{Hoferichter:2020lap,Zanke:2021wiq,Ludtke:2023hvz,Hoferichter:2024fsj}, and the matching to short-distance constraints~\cite{Bijnens:2019ghy,Bijnens:2020xnl,Colangelo:2019lpu,Colangelo:2021nkr}.

\section{Conclusions}
\label{sec:con}

We discussed some recent applications of dispersive approaches to the HVP and HLbL contributions to $(g-2)_\mu$. Besides consistency checks, we showed how detailed studies of the hadronic cross sections allow one to establish correlations with other low-energy observables, and could thereby help better understand tensions among experiments. Similarly, dispersive techniques can be applied to improve the calculation of radiative corrections and to estimate the size of isospin-breaking contributions, enabling a direct comparison to lattice QCD. For HLbL scattering, the dominant intermediate states have already been evaluated in a dispersive framework, and ongoing analyses of subleading contributions should allow one to match the precision goals defined by the Fermilab experiment.    

\section*{Acknowledgments}

Financial support by the SNSF (Project No.\ PCEFP2\_181117) is gratefully acknowledged.

\end{document}